# Machine Learning enabled models for YouTube Ranking Mechanism and Views Prediction

Vandit Gupta[1, a)*] and Akshit Diwan[1, b)] and Chaitanya Chadha [2, a)] and Dr. Ashish Khanna[1, c)] and Dr. Deepak Gupta [1, d)]

[1]*Maharaja Agrasen Institute of Technology, Plot No 1 Rohini, Plot No 1, CH Bhim Singh Nambardar Marg, Sector 22, PSP Area, Delhi, 110086, India*
[2]*SRM University, Haryana, 39, Rajiv Gandhi Education City, Sonipat, 131029, India*

*Author Emails:*
[a)] vanditgupta22@gmail.com
[b)]akshitdiwan05@gmail.com
[c)]chaitanyachadha12@gmail.com
[d)]ashishkhanna@mait.ac.in
[e)]deepakgupta@mait.ac.in

**Abstract.** With the continuous increase of internet usage in today's time, everyone is influenced by this source of the power of technology. Due to this, the rise of applications and games Is unstoppable. A major percentage of our population uses these applications for multiple purposes. These range from education, communication, news, entertainment, and many more. Out of this, the application that is making sure that the world stays in touch with each other and with current affairs is social media. Social media applications have seen a boom in the last 10 years with the introduction of smartphones and the internet being available at affordable prices. Applications like Twitch and Youtube are some of the best platforms for producing content and expressing their talent as well. It is the goal of every content creator to post the best and most reliable content so that they can gain recognition. It is important to know the methods of achieving popularity easily, which is what this paper proposes to bring to the spotlight. There should be certain parameters based on which the reach of content could be multiplied by a good factor. The proposed research work aims to identify and estimate the reach, popularity, and views of a YouTube video by using certain features using machine learning and AI techniques. A ranking system would also be used keeping the trending videos into consideration. This would eventually help the content creator know how authentic their content is and healthy competition to make better content before uploading the video on the platform will be ensured.

## INTRODUCTION

In today's advanced world of technology and resources, when the internet is connecting the entire world as one, the resources our generation access should have content of the highest quality. In this era of Youtube, Reddit, Facebook, and others, the content people watch on the largest technological platforms of the world, should be analyzed properly. Today, we have that same motivation to make the best user experience on the world's largest video-watching platform, YouTube. Youtube's audience is composed of different backgrounds, cultures, ages, and mindsets. To provide content that doesn't impose any emotional harm to any individual watching and is authentic is what every content creator shall have as their ultimate goal.

YouTube content creators shall be aware of the factors that depict the popularity of their videos so that they can improve their decision-making so that the next steps could be figured out. For example, a content creator might be working on a new video and have a list of title and description ideas, but without the help of tools for prediction and analytics, they have only their intuition to rely on when making a final decision on how good their video will have an influence. This aims to provide a way for content creators to rank their candidates in order of predicted view counts. Given a title, description, tag list, and the identifier of the content creator's channel, the prediction of views a video with the given data would get if it were posted.

Youtube Views Prediction is a way of enhancing the quality and authenticity of influencers who want to benefit from the biggest platform on the internet currently. Youtube has videos of many categories ranging from Music, Cooking, Gaming, etc. It is the second-most used social platform in the world, just running behind Facebook. Among such an enormous user base, it becomes tough to compete with other YouTubers' content. Many factors have to be considered to upload a video so that one can get the maximum views and therefore can benefit from continuous hard work. This paper aims at helping all those YouTubers to accurately evaluate their content without wasting many resources.

A ranking mechanism is also proposed in this paper which aids in giving a proper guided path for the maximum popularity of a video. The outreach is not only dependent on the content of the video, it is dependent on closeness with the trending topics within a specific time interval as well. As Youtube has a list of trending topics that is available to watch for the mass viewer base, if a relationship between the content creator's video and the top and trending topics of the platform could be obtained, the video would have a greater probability to achieve a specific target. The mechanism would be functional by using a framework called Word2Vec, which relies on Natural Language Processing. Its parameters would be as follows:
- Description and Tags of Video
- Trending Youtube topics

A matching score could be achieved by this technique to make the creator's experience even better, which eventually leads to amazing content on the platform.
It also ensures that in the world of the internet, the massive community of users on the biggest platform Youtube get the best content that they search for, given that the YouTubers have a healthy competition for providing authentic material which eventually benefits the content creators, motivating them to work in the right direction, as well as gives 2.3 billion people around the world the best experience

The main contributions of the paper are as follows:
- A web-based interactive tool to help predict the views on a Youtube video using certain parameters and algorithms.
- A proposed ranking mechanism to predict the improvement in rank and the reach of the video.

## RELATED WORKS

A lot of fine work has been done by previous researchers that have tried to explore analysis and prediction when it comes to social media platforms like Youtube. The dependence of views on parameters as mentioned in William Hoiles et al. [1] paper is of utmost importance and the selection of the features is important. Variables like the number of subscribers, description and language, and optimization of features like thumbnail and title, play a big role in prediction. Visualization is a key factor to analyse the data as one tries to work upon as studied by Quyu Kong et al. [2], let it be the correlation of features, frequency of uploads in a particular location, comparison of likes between two channels, etc. which eventually brings out the spread of a particular video. Feature selection as done by Yuping Li et al. [3] using PCA is an important step of the entire data preprocessing stage. Previously KNN and Stepwise Regression methods were applied by Soufiana Mekouar et al. [4] to find the best possible accuracy. Not only can views bring out the popularity of that video, rather they can bring out the scope of the future of that Youtube channel too. Moving one step ahead Ibrahim Said et al. [5] successfully applied regression algorithms to predict revenue that could be generated by reviewing the movie trailers. The success of a movie trailer is eventually linked to its box office collection. William Hoiles et al. [6] proposed a method of predicting the commenting behavior of youtube viewers using Reinforcement learning which is an important parameter in our paper. ChenYen-Liang et al. [7] worked on the early prediction of the popularity of video by creating a meta classifier that contained many base classifiers such as neural networks and Naive Bayes. Alireza Zohourian et al. [8] proposed a method to predict likes on Instagram posts using regression algorithms like SVM, Kernel-SVM, and Local Polynomial Regression. Adele Lu Jia et al. [9] predicted the views on the User Generated Content site which contained improved social features. Mathias Bärtl et al. [10] paper on a statistical analysis of youtube channels, uploads, and views was helpful in our Exploratory Data Analysis. Peter Braun et al. [11] proposed a method of using an outlier detection algorithm to reduce the noise in the dataset helping in improving the accuracy of the model. Changsha Ma et al. [12] introduced a new metric to predict the views on youtube early on by using the video's length. Kharkar et al. [13] also worked on the early growth of youtube views by using the history and used the past video's statistics to predict how well the current video would perform.

The pros and cons, along with the results of the previous works are presented in a tabular form in the following part below.

## Comparison between Related Work

**TABLE 1.** Pros and Cons along with results of previous research work

| Research paper | Pros | Cons | Results |
|---|---|---|---|
| William Hoiles et al. [1] (2017) | YoutubeRanking used better and more efficient parameters to ensure the target gets predicted with utmost accuracy | Description of the video is not considered which is an important parameter for predicting views on a video | RMSE: 0.47, $R^2$:0.80 |
| Yuping Li et al. [3] (2019) | Included description along with other major parameters in the model | The dataset contains imbalance which might lead to less accurate results | F1 score: 0.736 |
| Soufiana Mekouar et al. [4] (2017) | High accuracy model generated using logistic and stepwise regression | Better algorithms such as XGBoost and Random Forest are not tested. Stepwise regression also has some issues like providing biased results | Accuracy: Logistic regression: 91.80 % Stepwise regression: 91.83 % |
| Peter Braun et al. [11] (2017) | Successfully detected outliers in the youtube history of a user | Model proposed not achieving high accuracy | Overall support: 63% |
| Changsha Ma et al. [12] (2017) | Introduced a new alpha parameter to help predict long term views in the early stage of the video and mean absolute percentage error (MAPE) is less than the other state of the art models | The Lifetime Awareness Regression model proposed in the paper has some performance bottlenecks due to the lack of more advanced features | Mean absolute percentage error (MAPE) - 0.2773 |

| Kharkar [13] et al. (2020) | Predicted views with as few parameters as possible with considerable accuracy | 66 % accuracy of the model | Accuracy: 66 % |

## Parameters Used

Many researchers have predicted views on YouTube videos before by using various parameters. The parameters used by them and by our model are mentioned in the image below.

The abbreviations used in Table2 are as follows:
- VID - Video ID
- Title - Video Title
- Age - Difference between the Publishing date and Trending date
- EV - Early Views ( Used for Early prediction of views in some papers)
- CT - Channel Title
- CG - Category of the video
- LV - Length of Video
- TG - Tags used by the uploader
- LK - Likes on the video
- DK - Dislikes on the video
- CC - Comment Count
- TN - Thumbnail of the video
- DS - Description of the video

**TABLE 2.** Pros and Cons along with results of previous research work

| Paper | VID | Title | Age | EV | CT | CG | LV | TG | LK | DK | CC | TN | SC | DS |
|---|---|---|---|---|---|---|---|---|---|---|---|---|---|---|
| W.Hoiles et al. [1] (2017) | | ✓ | | | | | | ✓ | | | | ✓ | | |
| Y.Li [3] et al. (2019) | | ✓ | ✓ | | | | ✓ | ✓ | | | | | | ✓ |
| S.Mekouar [4] et al. (2017) | ✓ | | ✓ | ✓ | ✓ | ✓ | ✓ | | | | ✓ | | | |
| C. Ma [12] et al. (2017) | | | ✓ | | ✓ | ✓ | ✓ | | ✓ | ✓ | ✓ | | ✓ | |

| | | | | | | | | | | |
|---|---|---|---|---|---|---|---|---|---|---|
| Kharkar [13] et al. (2020 | ✓ | | | ✓ | | | | ✓ | | |
| V. Gupta ( Proposed Model ) et al. | ✓ | ✓ | | ✓ | ✓ | ✓ | ✓ | ✓ | ✓ | ✓ |

# PROPOSED METHODOLOGY

To predict a view count for any given video, a model has to be trained using parameters of a dataset that are related to the target variable i.e, views. This is a regression problem as the target variable is numerical and continuous in nature. A suitable model has to be chosen with feature selection and an in-depth data analysis has to be implemented. So, a step-by-step process has to be undertaken from gathering the dataset to obtaining the view prediction of any random video to generate the ranking of that video. The first step in the process is to import the dataset on which data analysis and modeling would be implemented.

As depicted by the flowchart (Figure 1) (below), the first step would be to gather the dataset which contains specific data required for training the model. The next step would be to perform Exploratory data analysis to find the outliers, feature correlation, and gain other useful insights. After that, the dataset is modified so that the time difference between the date of upload and the date at the time of prediction can be calculated and used as parameters to improve the model. After dataset preparation, the model is trained on IBM Watson using specific regression algorithms. An API key is generated which is used for integration with the User interface. Finally, the output is generated and visible to the user through the user interface.

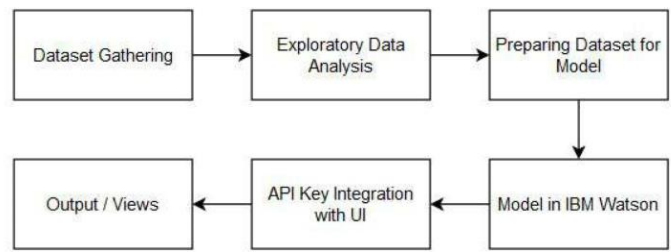

**FIGURE 1.** The working flow of the model

All the above steps are discussed further in detail in section 3.

## Dataset Gathering

We used the 'YouTube Trending Video Dataset' from Kaggle which can be accessed from 'https://www.kaggle.com/rsrishav/youtube-trending-video-dataset' and selected the trending videos of India and the US from the files 'IN_youtube_trending_data.csv' and 'US_youtube_trending_data'. Both datasets contain 1048576 unique videos at the time of this paper. The Training data and Testing data have been split into 70:30 ratios.

The dataset has these defined parameters:

**TABLE 3.** Parameters in the dataset

| Parameter | Meaning |
| --- | --- |
| Video_id | ID associated with the video |
| Title | Title of the video |
| publishedAt | Time at which video was uploaded |
| channelId | ID associated with the youtube channel |
| channelTitle | Channel name |
| categoryId | Category to which video belongs to |
| Trending_date | Trending video dates / Date up to which prediction has to be done |
| Tags | Tags associated with video |
| View_count | View count to be predicted / Target Variable |
| Likes | Number of likes at the time of upload |
| Dislikes | Number of dislikes at time of upload |
| Comment_count | Number of comments at the time of upload |
| Thumbnail_link | Image useful for NSFW and Clickbait detection |
| Comments_disabled | Are comments disabled or not |
| Ratings_disabled | Are ratings disabled or not |
| Description | Description of the video |
| Country | Country of upload |
| CountryId | The ID of the country of upload |

## Exploratory Data Analysis (EDA)

Exploratory Data Analysis along with a comparative analysis of both the datasets is implemented. The steps of the EDA are explained further below.

### Feature Correlation

A correlation matrix was first generated to help us know which parameters were significant in predicting views and we filter out the rest of the parameters to make the final model more accurate. To visualize the correlation matrix heatmap was created.

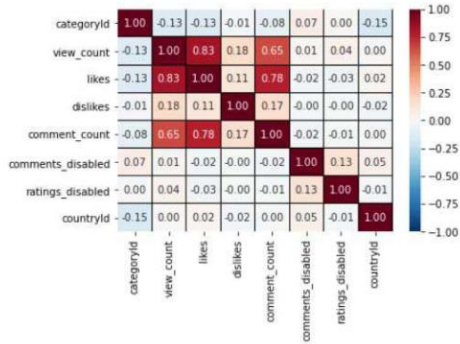

**FIGURE 2.** Heatmap for Correlation between features

0.5 Threshold was set for the heatmap. Figure 2 suggested that View_count and comments were the significant parameters to predict likes on the video.

*Outlier Analysis*

After that outlier analysis was conducted to find those videos which were not following the general trend and could cause the model to deviate. Those outliers were detected and removed. To visualize outliers, a Box plot was used as shown in Figure 3

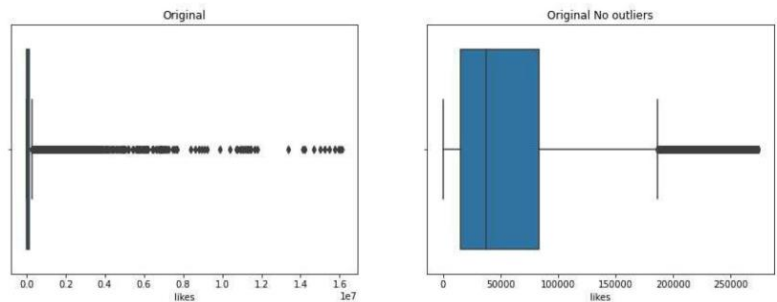

**FIGURE 3.** Outlier Detection

# Data Preparation considering Time Period

For preparing the dataset for the model, we also considered the time difference to predict views in a specific time interval. For that, we had to modify our dataset and calculate the difference between the date of upload and the date up to which we wanted the results. For calculating that, we formatted the 'publishedAt' and 'trending_date' columns and split them into differences between dates, hours, and months.

This splitting into columns helped to calculate the time difference so that the regression model could also carry out finding a pattern between the time difference of the upload time and any given time, and the view count.

The Abbreviations used in Table 4 are as follows:
- CID - Category ID
- Views - Views on the video
- CD - Are Comments disabled or not
- RD - Are Ratings disabled or not

- DD - Difference in days between trending date and date of upload
- DH - Difference in hours between trending date and date of upload
- PY - Year of publishing of video
- PM - Month of publishing of video
- TY - Trending Year
- TM - Trending Month

**TABLE 4.** Updated dataset post adding time difference for the videos

| CID | Views | CD | RD | DD | DH | PY | PM | TY | TM |
|---|---|---|---|---|---|---|---|---|---|
| 24 | 834299 | False | False | 3 | 85 | 2018 | 4 | 2018 | 4 |
| 1 | 61240 | False | False | 2 | 50 | 2018 | 1 | 2018 | 1 |
| 19 | 573049 | False | False | 7 | 178 | 2018 | 4 | 2018 | 4 |
| 10 | 16408326 | False | False | 15 | 366 | 2018 | 4 | 2018 | 5 |
| 26 | 2491725 | False | False | 11 | 286 | 2018 | 5 | 2018 | 5 |

## Algorithm used

A regression algorithm would be used to solve this problem. Several types of Regression can be applied to get the best accuracy and boosting models but the best results were obtained on **XGBoost, Decision Tree Regressor, or Random Forest Regressor.**

**XGBoost:** XGBoost is a boosting algorithm aiming to create a strong regressor tree based on working and updating weak regression trees. It is an ensemble method.

**Decision Tree:** A decision tree is a regression algorithm that is built top-down from a root node and involves splitting the data into further subtrees based on the selection of metrics which can be either information gain or Gini index. The main goal is to reduce the entropy and split on that parameter that provides the maximum information gain. At each node of the tree, a test is applied to the predictor variables and depending on the value of the outcome of the test we either go to the left or right subtree. Training data is used to construct the decision tree and testing is used for predicting the value through the tree constructed by training data

**Random Forest**: Random forest is an ensemble algorithm, which means it is a collection of many decision trees that are each trained on random sub-samples of the training set. The decision trees in the random forest use different parameters from each other. The final prediction of the Random Forest is given as the mean prediction of the individual decision trees.

Before proceeding to the model, post-EDA, the main visible predictors were as follows:

- Channel Name
- Video Title
- Type of video (Music, Cooking, Cars, Reaction Video) / Category ID
- Number of Comments
- Number of tags used
- Tags present alongside Title
- Description
- Publishing Date of Video
- Trending Date of Video
- Likes
- Dislikes

The integral element of any video which is visible is the title. Using the title, some references can be observed as written below:

The ranking mechanism can be enhanced by taking the title into NLP(Natural Language Processing) networks such as Word2Vec or Glove for recommendation systems. This can be matched with the top 100 topics on Youtube in a particular week through Google trends. Videos with titles matching the Trending videos would receive a boost in ranking which would eventually increase the views as well. The mechanism has been discussed further in the paper later.

## IBM Watson Model

Our model has to take multiple parameters and uses regression to predict views of a particular video using its parameters like the number of Likes, Dislikes, Title, Description, Tags, etc.

IBM Watson is a cloud service that takes the dataset, preprocesses it, and tries to use a different variety of algorithms to predict the outcome generating the best accuracy possible for the given problem. The Auto ML service of IBM Watson was used for searching for the best possible results.
We provided IBM Watson with the processed dataset and applied the regression algorithms to predict the views. The model was deployed using IBM Watson and an API Key was generated. This key is used to integrate the model with the user interface.

Finally, the model is user-ready to take inputs and present the views coming directly from the model.
Frameworks for Web Development were used:
- HTML
- CSS
- Javascript
- NodeJS

Now the video data will be passed for a ranking mechanism which will be done through Natural Language Processing.

## Natural Language Processing

Natural Language Processing is the branch of Machine Learning which works with text. As the description and tags are a very integral part of any Youtube video, we could use them to extract some more meaning out of it, not just limiting their use to predict the user's views. We would be implementing Word2Vec for using word embeddings to develop matching scenarios for tags in descriptions with trending topics.

The entire process is explained in the next section.

## Ranking System

To improve the ranking system, the description and tags can be used.

### *Word2Vec*

Word2Vec is a neural network-based model used in recommendation systems when working with text. It combines deep learning and Natural Language Processing. It has one neural layer and has previously trained embeddings. Word embeddings are representations of a variety of words that can be used to check the similarities and differences among words. Word2Vec provides pre-trained embeddings which can be manipulated further by training further on a new set of data.

Pre-trained embeddings can be accessed in two modes:
1, Static: The embedding doesn't get updated with the training of the model.

2. Dynamic: The embedding gets updated during the training of the model

Further, Word2Vec classification can be done based on the type of prediction of target:
Word2Vec is of two types mainly:
- CBOW
- Skip Gram Model

Skip Gram is used when recommendations have to be calculated by a single word, whereas CBOW is in reverse, i.e, multiple words give one word as a recommendation. This paper will use static embeddings, based on Google News. When the entire embedding is imported, any given two words can be checked for closeness in meaning using a metric known as cosine similarity. Cosine Similarity is a measure of similarity of any two given vectors in three-dimensional space. The greater the value, the closer in meaning the vectors associated, and so are the words concerned with the vectors respectively. Cosine Similarity plays a major role in finding recommendations using the embeddings in the model to give some useful insights.

With enough data and context, Word2vec can make guesses about a word's meaning based on how many times it has appeared in the corpus before. The guesses can help to make the corpus so that the distance between them makes the equation of similarity and difference clear. Man to Woman relationship is similar to King to Queen relationship. So the link between two words can be delivered by this process. These can form the basis of sentiment analysis, and recommendation systems for many useful applications in diverse fields like healthcare, e-commerce, and research.

The output of the Word2Vec model is a corpus or a vocabulary that represents words as vectors so that cosine similarity could be calculated and could be used further for textual analysis and for serving multiple purposes in Natural Language Processing.

A matching algorithm can be applied to description tags that can be related to the top 100 topics of the month. A pre-trained version of Word2Vec will be used as the matching algorithm that can be tried for achieving the best accuracy. It will be trained on the Google News dataset which helps to find the similarity as shown below.

First and foremost, trending data has to be gathered. For achieving this, Google Trends can be used. Google Trends is a website that displays information about the most searched topics in the time period specified. As Youtube belongs to Google, it can show the current information as shown below.

From Google Trends, data could easily be extracted and it could be checked for similarity with the tags and the keywords in the description of a video. If the matching value is high, the ranking and success rate increase. The pre-trained Word2Vec will help out finding similarity scores.

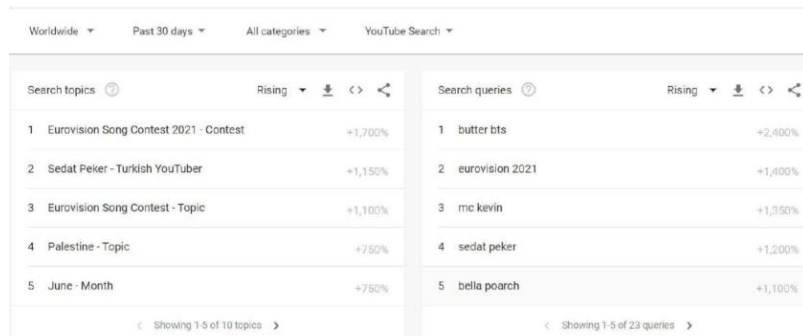

**FIGURE 4.** World Trending Topics of Youtube for the past month

As can be seen in Figure 4, the top topic is Eurovision, which is a singing contest. It is passed to the Word2Vec model.

## RESULTS AND DISCUSSIONS

After the integration is complete, we finally test on random videos. Figure 5 shows the User Interface for the content creator to input the important details for the prediction of views of the video.

**FIGURE 5.** UI for the content creator to enter details about the video

Figure 6 shows the number of predicted views for the video.

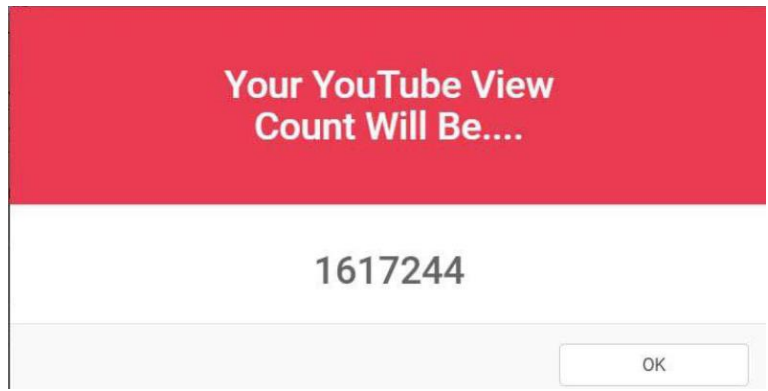

**FIGURE 6.** Prediction of Views

Figure 7 represents the comparison between predicted views and actual views on real videos uploaded on Youtube.

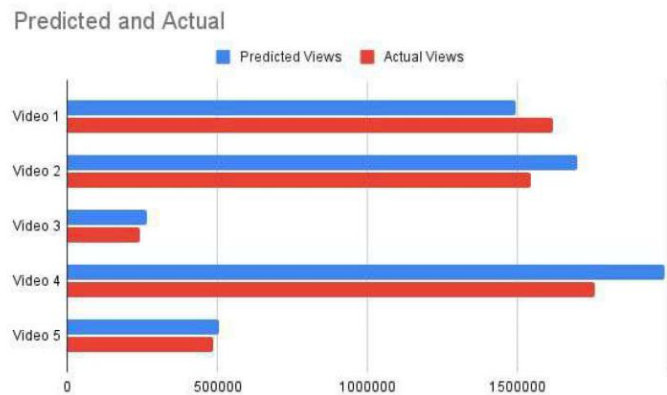

**FIGURE 7.** Accuracy analysis for random testing data with different videos on youtube

**TABLE 5.** Predictive Models and their Training and Testing scores

| Model | Training Accuracy (%) | Testing Accuracy (%) |
|---|---|---|
| XGB Classifier | 87.8 | 75.2 |
| Decision Tree | 82.5 | 70.9 |
| Random Forest | 89.6 | 78.4 |

The Youtube View Prediction platform is successfully able to provide results due to algorithms used with a well-defined dataset that we used. Deep analysis of the data also helped to gain insights into the category and genre of the video people prefer from different diversities and backgrounds.

# MANAGERIAL AND SOCIAL IMPLICATIONS

This paper proposes an improvised digital experience for the massive user base on Youtube. People from all cultures watch the streaming medium by searching for songs, news, movies, study material, dancing videos, piano, and many more categories. The number of categories can't be defined, so the diversity is too broad. So, it becomes necessary for content developers to provide the best version of their work to the world. This will ensure a good future for the creator, as well as the audience, who can enjoy and learn from the platform which is why it exists today, to make people comfortable with the material it has. Using Views Prediction and Ranking Mechanism, the best content can be selected and published so that the chances of talent and creativity reaching the user base increases, which ensures the best possible scenarios for both the users as well as the content creator.

# CONCLUSION AND FUTURE WORK

In this paper, we performed Exploratory Data Analysis on the YouTube dataset to gain meaningful insights as well as create a model to predict views on a video based on certain parameters to help content creators predict the reach of their video. This would help the content creators to create better content for the platform. We have also proposed a ranking mechanism to help predict if the video would be trending by comparing the keywords of the video with the top trending keywords retrieved by Google Trends. The dataset also contains thumbnails. In the future, thumbnails can also be used for predicting views. For example: If someone uploads a Youtube Video having Donald Trump in the thumbnail at the time of the US elections, it is bound to have higher views than any random thumbnail. Moreover, as people from all age groups are a part of the Youtube audience, the thumbnail can be checked for inappropriateness (NSFW Detection). Convolutional Neural Networks can be used to check the thumbnail for inappropriateness. The thumbnail can be passed through the PyTesseract OCR model to detect text, which will further be checked with the top trending topics of that month on Youtube. If there is a similarity, the predicted ranking of the video will get boosted.

# REFERENCES

1. William Hoiles, Anup Aprem, Vikram Krishnamurthy, "Engagement and Popularity Dynamics of YouTube Videos and Sensitivity to Meta-Data " in *IEEE TRANSACTIONS ON KNOWLEDGE AND DATA ENGINEERING* Volume: 29, Issue: 7, 1st July 2017, pp.1426 - 1437
2. Quyu Kong, Marian-Andrei Rizoiu, Siqi Wu, and Lexing Xie. 2018, "Will This Video Go Viral? Explaining and Predicting the Popularity of Youtube Videos." in *WWW '18 Companion: The 2018 Web Conference Companion*, April 23–27, 2018, Lyon, France. ACM, New York, NY, USA, pp. 175–178
3. Yuping Li, Kent Eng, Liqian Zhang "YouTube Videos Prediction: Will this video be popular?" in *Civil and Environmental Engineering*, Stanford 2019, pp. 1-6

4. Soufana Mekouar, Nabila Zrira and El-Houssine Bouyakhf "Popularity Prediction of Videos in YouTube as Case Study: A Regression Analysis Study" in *Proceedings of the 2nd International Conference on Big Data, Cloud and Applications*, March 2017, Article No.: 51, pp. 1–6
5. Ibrahim Said, Ahmad, Azuraliza, Abu Bakar and Mohd Ridzwan Yaakub "Movie Revenue Prediction Based on Purchase Intention Mining Using YouTube Trailer Reviews" in *Information Processing & Management* Volume 57, Issue 5, September 2020
6. William Hoiles, Vikram Krishnamurthy and Kunal Pattanayak "Rationally Inattentive Inverse Reinforcement Learning Explains YouTube Commenting Behavior" in *Journal of Machine Learning Research*, Volume 21,24th September 2019, pp. 1-39
7. ChenYen-Liang and ChangChia-Ling, "Early prediction of the future popularity of uploaded videos" in *Expert Systems with Applications* Volume 133, 1 November 2019, pp. 59-74
8. Alireza Zohourian, Hedieh Sajedi, and Arefeh Yavary, "Popularity prediction of images and videos on Instagram" in *2018 4th International Conference on Web Research (ICWR)*, 2018, pp. 111-117
9. Adele Lu Jia, Siqi Shen, Dongsheng Lib, and Shengling Chen "Predicting the implicit and the explicit video popularity in a User Generated Content site with enhanced social features." in *Computer Networks Volume* 140, 20 July 2018, pp. 112-125
10. Mathias Bärt,l "YouTube channels, uploads, and views: A statistical analysis of the past 10 years" in *Convergence: The International Journal of Research into New Media Technologies*, Volume 24, Issue 1, pp. 16–32
11. Peter Braun, Alfredo Cuzzocrea, Lam M.V. Doan, Suyoung Kim, Carson K.Leung, Jose Francisco A. Matundan and Rashpal Robby Singh, "Enhanced Prediction of User-Preferred YouTube Videos Based on Cleaned Viewing Pattern History" in *Procedia Computer Science* Volume 112, 2017, pp. 2230-2239
12. Changsha Ma, Zhisheng Yan, and Chang Wen Chen, "LARM: A Lifetime Aware Regression Model for Predicting YouTube Video Popularity" in *Proceedings of the 2017 ACM on Conference on Information and Knowledge Management*, November 2017, pp. 467–476
13. Kharkar and Ritvik "Predicting and Characterizing Early Growth of YouTube Videos" Ph.D. thesis in *Open Access Publications from the University of California* (2020), UCLA, 2020